\documentclass[reprint,superscriptaddress,showpacs,amsmath,amssymb,aps,prl]{revtex4-1}
\usepackage{graphicx}  
\usepackage{dcolumn}   
\usepackage{bm}        
\usepackage{amssymb}   
\usepackage{mathrsfs} 

\usepackage[colorlinks=True,urlcolor=blue,linkcolor=blue,citecolor=blue]{hyperref}
\usepackage{xcolor}
\usepackage{multirow}
\usepackage{setspace}
\usepackage{CJK}
\usepackage{color,soul}

\begin{document}



\begin{CJK*}{UTF8}{}

\title{Generic Symmetry Breaking Instability of Topological Insulators due to a Novel van Hove Singularity}

\date{\today}
\author{Xu-Gang He (\CJKfamily{gbsn}何绪纲)}
\affiliation{Condensed Matter Physics and Materials Science Department,
Brookhaven National Laboratory, Upton, New York 11973, USA}
\affiliation{Department of Physics and Astronomy, Stony Brook University, Stony Brook, New York 11794, USA}
\author{Xiaoxiang Xi (\CJKfamily{gbsn}奚啸翔)}
\affiliation{Photon Sciences,
Brookhaven National Laboratory, Upton, New York 11973, USA}
\author{Wei Ku (\CJKfamily{bsmi}顧威)}
\affiliation{Condensed Matter Physics and Materials Science Department,
Brookhaven National Laboratory, Upton, New York 11973, USA}
\affiliation{Department of Physics and Astronomy, Stony Brook University, Stony Brook, New York 11794, USA}

\begin{abstract}
We point out that in the deep band-inverted state, topological insulators are generically vulnerable against symmetry breaking instability, due to a divergently large density of states of 1D-like exponent near the chemical potential.
This feature at the band edge is associated with a novel van Hove singularity resulting from the development of a Mexican-hat band dispersion.
We demonstrate this generic behavior via prototypical 2D and 3D models.
This realization not only explains the existing experimental observations of additional phases, but also suggests a route to activate additional functionalities to topological insulators via ordering, particularly for the long-sought topological superconductivities.
\end{abstract}

\maketitle
\end{CJK*}

Topological insulators have attracted great research interests recently, due to their realization of topological characteristics in fundamental physics and their significant potential in spintronic applications~\cite{M2010,Z2006,XZ2011,KM2005, Li2007,li2007, Li2011}.
For example, the edge states are spin-momentum locked and can propagate without dissipation, enabling exciting novel applications in magnetic transistors~\cite{B2010}, spin battery~\cite{Ma2010}, and magnetoresistive devices~\cite{Chu2009}.

Interestingly, topological insulators are often found to be accompanied with additional symmetry breaking phases.
For examples, Cr-doped Bi$_2$(Se$_x$Te$_{1-x}$)$_3$ is found to enter a magnetic phase~\cite{zj2013}, allowing the realization of the long-sought quantum anomalous Hall effect~\cite{chang2013}.
Another example, the charge density wave instability by the chiral symmetry breaking in the 3D Weyl semimetals, made by the topological insulator multilayer~\cite{Bur2011}, has been proposed to form axion insulators, with the dissipationless transport on the axion strings~\cite{Wang2013}.
One thus wonders ``Is there a generic reason for the strong tendency toward symmetry-breaking instability in the topological insulating phase?" and ``How should one apply it to realize new quantum phenomena and to tailor their unique functionalities?" 

Perhaps the most exciting possibility is to realize the topological superconductor~\cite{Read2000, XZ2011, Kit2009, liang2010, Qi2010, Roy2008,Qi2009}.
The edge state of a topological superconductor has a peculiar nature that it is its own anti-particle, a special particle named Majorana fermion~\cite{Read2000}.
The Majorana fermion has some exotic properties that make them scientifically interesting, such as the non-Abelian statistic rather than the Bose-Einstein statistic of the bosons or Fermi-Dirac statistic of the normal fermions~\cite{Wil2009}.
This also allows them to be used for practical applications, such as to create Majorana qubits and to realize the topological quantum computation~\cite{Bravyi2006, Stern2010, Leijnse2012}.
So far, the main thinking of the field is to utilize the superconducting proximity effect to create topological superconducting state at the interface between a topological insulator and a fully gapped superconductor~\cite{Fu2008}.
However, it would be highly desirable to also explore the intrinsic \emph{bulk} superconducting instability of doped topological insulators.

In this letter, we point out a generic feature in the topological insulating phase that renders the electronic system vulnerable against symmetry-breaking instabilities.
Deep into the topological phase, the inverted band unavoidably develops a Mexican-hat dispersion that gives rise to a novel van Hove singularity (VHS)~\cite{van1953} at the band edge in both 2D and 3D.
In essence, the geometry of the Mexican-hat dispersion hosts a singular density of states (DOS) with a 1D-like divergent exponent.
In doped systems, this may also cause a Lifshitz transition--a change of Fermi surface topology, involving appearance of additional disconnected Fermi sheets with characteristic shapes.
In the absence of Fermi surface nesting, the divergent DOS would particularly enlarge the phase space of the Cooper pairing channel and favor the superconductivity.
These generic features, which we will demonstrate with prototypical 2D and 3D models, not only explains the observed broken symmetry states in many topological systems, but also suggest a clear route to activate additional functionalities, particularly the long-sought topological superconductivity.

\begin{figure*}
\includegraphics[height=9.4cm,width=18cm]{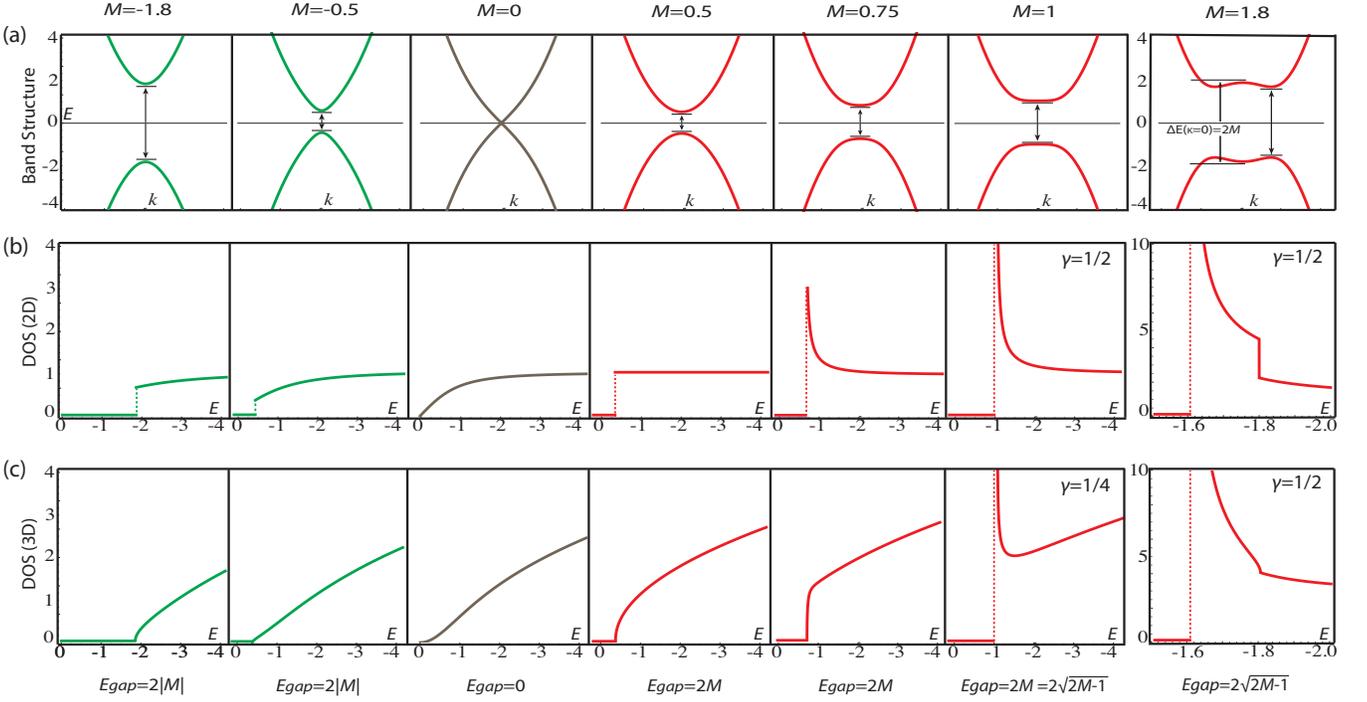}
\caption{\label{fig:fig1}
(color online; arb. unit.)  (a) Evolution of the band dispersion according to the Eq.(~\ref{eq:eqn3}), (b) The change of the DOS for 2D system and (c) for 3D system, from a normal insulator ($M<0$) to a topological insulator ($M>0$).
Each column is labelled by the mass term at the top, and the length of the double-headed arrows describe the band gaps given at the bottom.
When the band demonstrates Mexican-hat dispersion at $M>1$, the DOS diverges at the band edge $\sim |\omega|^{ - \gamma}$ with a 1D-like divergent exponent $\gamma=1/2$. 
The regular VHS corresponding to the tip of the Mexican hat (an additional step function in 2D and square root function in 3D), can also be observed at $|\omega|=1.8$ in the right most panels.
}
\end{figure*}

\begin{table}[h]
\advance\leftskip-0.3cm
\begin{tabular}{|c||c|c|c|c|}
\hline
\hline
 {$g\left( \omega \right)$} & $M \ll 0$ &   $  =0 $ &   $ =1$ &  $>1$ \\
\hline
 2D & $\frac{1}{\pi }\frac{{\left| M \right|}}{{1 - M}} + \frac{1}{\pi }\frac{{1 - 2M}}{{{{\left( {1 - M} \right)}^3}}}|\omega| $   & $\frac{|\omega| }{\pi }$ & $\frac{1}{{2\pi }}{\left( {\frac{{{E_{gap}}}}{{|\omega|}}} \right)^{\frac{1}{2}}}$ &$ \frac{1}{{  \pi }}{\left( {\frac{{{E_{gap}}}}{{|\omega|}}} \right)^{\frac{1}{2}}}$  \\
\hline
3D &  {$\frac{{\sqrt 2 }}{{{\pi ^2}}}{\left( {\frac{{\left| M \right|}}{{1 - M}}} \right)^{\frac{3}{2}}}\sqrt {|\omega|}  $ }  & $\frac{{{\omega ^2}}}{{{\pi ^2}}}$ & $\frac{1}{{{\pi ^2}}}{\left( {\frac{{{E_{gap}}}}{{|\omega|}}} \right)^{\frac{1}{4}}} $ & $ \frac{K_M}{{ \pi ^2 }}{\left( {\frac{{{E_{gap}}}}{{|\omega|}}} \right)^{\frac{1}{2}}}$ \\
\hline
\end{tabular}
\caption{\label{tab:tab1}
Different analytical limit of the DOS at the band edge from a normal insulator ($M<0$) to a topological insulator ($M>0$).
With a Mexican hat band, the DOS diverges at the band edge with a 1D-like exponent, for both 2D and 3D systems.
The 3D DOS is also proportional to the radius--${K_M}$ of the bottom sphere of the Mexican hat dispersion (cf. Fig.~\ref{fig:fig2}).}
\end{table}

We start by examining the evolution of the electronic bands structure across the topological phase transition via a band inversion.
For a 2D system, we use a sub-block of the BHZ model that drops the spin degree of freedom~\cite{Bernevig2006,Ph2012}.
It represents the simplest topological insulator (the Chern insulator) with the bands inversion scenario:
\begin{equation}
\label{eq:eqn1}
H(k) = d_a (k){\sigma ^a},
\end{equation}
where the 3-dimensional vector $\vec d$ is defined as $d_x (k) = \sin {k_x } $, $d_y (k) = \sin {k_y } $, $d_z  =  2 - M - \cos {k_x } - \cos {k_y } $, and $\sigma^a$'s are the Pauli matrices.
Note that all models of Chern insulators can be reduced to the similar form~\cite{Ph2012}.
The dispersions of this two-band system are $E_ \pm   =  \pm \sqrt {|\vec d|^2 }$.
Near the band inversion point (set as $|k|=0$), the Hamiltonian is, up to the 2nd order of $k'$s,
\begin{equation}
{H^{\left( 2 \right)}_{2D}}\left( k \right) = \left( {\begin{array}{*{20}{c}}
{\frac{1}{2}k_x^2 + \frac{1}{2}k_y^2 - M}&{{k_x} - i{k_y}}\\
{{k_x} + i{k_y}}&{-(\frac{1}{2}k_x^2 + \frac{1}{2}k_y^2 - M)}
\end{array}} \right);
\end{equation}
with the dispersions:
\begin{equation}
\label{eq:eqn3}
E_ {\pm} ^{\left( 2 \right)} \left( k \right) =  \pm \sqrt {k^2  + {{\left( {\frac{1}{2}k^2 - M} \right)}^2}},
\end{equation} 
where $k^2 = k_x^2 + k_y^2$.

For generic 3D cases, we use the simplified model of the topological insulator family including such as ${\rm{B}}{{\rm{i}}_2}{\rm{S}}{{\rm{e}}_3}$ crystal, which had been realized in the experiments~\cite{Zhang2009, Ch2009}:
\footnotesize{
\begin{equation}
H_{3D}^{\left( 2 \right)}\left( k \right) = \left( {\begin{array}{*{20}{c}}
{M - \frac{1}{2}{k^2}}&{{k_z}}&0&{{k_ - }}\\
{{k_z}}&{ - (M - \frac{1}{2}{k^2})}&{{k_ - }}&0\\
0&{{k_ + }}&{M - \frac{1}{2}{k^2}}&{ - {k_z}}\\
{{k_ + }}&0&{ - {k_z}}&{ - \left( {M - \frac{1}{2}{k^2}} \right)}
\end{array}} \right),
\end{equation}
\normalsize{
where ${k_ \pm } = {k_x} \pm i{k_y}$.
The resulting dispersions have the same form as Eq.(~\ref{eq:eqn3}), but with ${k^2} = k_x^2 + k_y^2 + k_z^2$.
}

\begin{figure}
\advance\leftskip-1.8cm
\includegraphics[height=5.8cm,width=7cm]{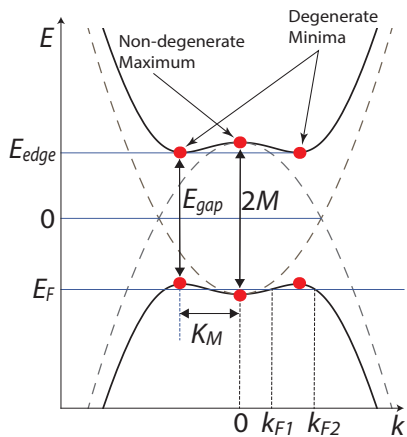}
\caption{\label{fig:fig2}
(color online) Formation of the Mexican-hat dispersion deep in the band-inverted state.
Dashed lines shows the inverted bands without the inter-band coupling.
Introduction of the inter-band coupling makes the system insulating via a gap opening.
As long as the band-inversion is stronger than the gap opening, a Mexican-hat dispersion is unavoidable.
$K_M$ denotes the radius of the bottom of the Mexican hat.}
\end{figure}

Figure~\ref{fig:fig1} summarizes the evolution of the band structure and the corresponding DOS's, from the topologically trivial phase ($M < 0$) to the topologically non-trivial phase ($M > 0$) in both 2D and 3D.
As expected, at the phase boundary ($M=0$) a metallic state is guaranteed~\cite{Z2006,Li2007,li2007}. 
Near the Dirac point, where two bands coincide in energy, the dispersion is linear and the DOS approaches zero.

Notice that deep into the topological phase ($M>1$), the system develops a Mexican-hat band dispersion.
This development of band structure is easily understood from Fig.~\ref{fig:fig2}.
When the band inversion is stronger than the gap opening between the two bands ($2M>E_{gap}=2\cdot min\{|E(k)|\}$), the dispersion unavoidably evolves into a Mexican hat.
Obviously, the development of such a feature is generic in all band-inversion scenarios and has in fact been observed in various numerical studies~\cite{Zhang2009, Y2012, Si2014}

The appearance of the Mexican-hat dispersion has important physical consequences.
For example, the DOS,
\begin{equation}
\label{eq:eqn2}
{g_D}\left( \omega  \right) = 2\int {\frac{{{d^D}k}}{{{{\left( {2\pi } \right)}^D}}}\delta \left( {|{E_{D,\pm}}\left( k \right)| - \left( {|\omega | + \frac{{{E_{gap}}}}{2}} \right)} \right)} 
\end{equation}
becomes divergent at the band edge.
Here, $\omega$ is the energy measured from the band edge, the factor 2 accounts for the spin degree of freedom, and $D=2,3$ for 2D and 3D cases, respectively.
Indeed, Fig.~\ref{fig:fig1}(b)(c) and TABLE~\ref{tab:tab1} show that the DOS's for both 2D and 3D topological systems diverge at the Mexican hat band edges with the divergent exponent, $\gamma=1/2$.
This divergent rate of the DOS is the same one found at the band edge of a regular1D system.
Actually, these 1D-like divergent DOS's are consistent with several recent experimental observations~\cite{A2012,P2012,Xiao2014}.

Such singular DOS's suggest that deep into the topological insulating phase, systems with band inversion are intrinsically vulnerable against symmetry breaking instabilities.
This is because of the proximity of the chemical potential to the DOS singularity due to the smallness of the band gap or the intrinsic doping of the systems.
For systems with near nested band structure, this may lead to charge density wave or spin density wave states.
Otherwise, more generally, this would enhance the instability in the $q=0$ channel, such as ferromagnetism or superconductivity, since their bare susceptibility are proportional to the DOS at the chemical potential~\cite{comment}.
This offers the most natural explanation of the recent observation of additional superconducting instability in the topological phase~\cite{Chao2011,JLZhang}.

\begin{figure}
\includegraphics[height=3.8cm,width=9.2cm]{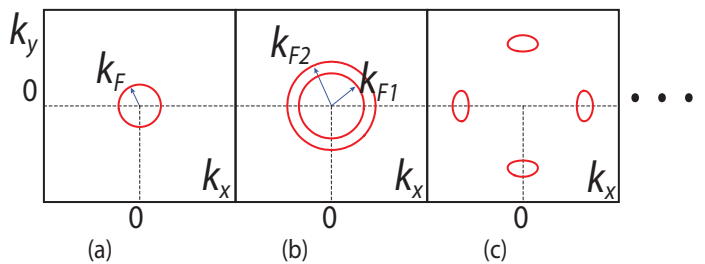}
\advance\leftskip-0.4cm
\caption{\label{fig:fig3}
(color online) Demonstration of the Lifshitz transition occurring in doped topological insulators: a single sheet of Fermi surface (a) would turn into two sheets of Fermi surface (b) due to formation of the Mexican-hat dispersion(cf. Fig.~\ref{fig:fig2}).
Inclusion of strong anisotropy might even split the Fermi surface into more sheets (c).}
\end{figure}

Compared to the recent proposal of interface superconductivity in topological crystalline insulators with flat bands~\cite{E2014}, the bulk superconductivity created via the Mexican hat in doped topological insulators has its advantage.
While it does suffer from low carrier density, the potential drawback of smaller phase stiffness~\cite{D1990, Emery1994,Emery1995}, should be compensated by the large kinetic energy (the relative big band width) and the bulk nature of the superconductivity.
In essence, in terms of the superfluid behavior, it would be in the same regime as the underdoped high-$T_c$ cuprates.

Fundamentally, notice that the VHS created by the Mexican hat is a qualitatively new class of VHS on its own, different from the known ones~\cite{van1953}.
Originally, based on the Morse theory and the quadratic dispersion of the momenta, the singular DOS is characterized by the non-degenerate extremum or saddle point~\cite{van1953,Ba1975}, which at most, can lead to a kink in 3D or a logarithmic divergence in 2D in general.
Only in the rare case of the flat bands (here as $M=1$), it is possible to realize more singular DOS caused by the quartic dispersion.
On the other hand, the Mexican-hat dispersion hosts the degenerate extrema at the bottom of the hat band, in addition to the common non-degenerate extremum at the tip of the Mexican hat.
The former gives rise to the 1D-like divergent behavior of the DOS and the latter produces the regular VHS, an additional step function in 2D and square root function in 3D (cf. the right most panels of Fig.~\ref{fig:fig1}).
Although at the band edge, the dispersion relations can still be approximated by the quadratic momenta, the degenerate extrema (a ring in 2D and a sphere in 3D) have 1D codimension, and consequently the DOS diverges at the band edge, like a 1D system~\cite{Ba1975}.

Interestingly, the appearance of the Mexican hat dispersion may also give rise to a Lifshitz transition in a doped system.
For example, observing the Fermi surface evolution of a doped systems in Fig.~\ref{fig:fig2} and~\ref{fig:fig3} (a)$\&$(b), one finds that the number of Fermi surface grows from one to two per Dirac point.
Across the Lifshitz transition, the non-analytical change of the corresponding DOS is known to lead to salient effects on thermodynamic, transport or magnetic properties~\cite{Bl1994}.
One thus expects clear signatures of such a transition in most measurements.
This could also be another way to qualitatively understand the enhancement of superconductivity discussed above.

It might also be instructive to make a connection to the similar singular DOS at the gap edge of a fully gapped superconductor with weak gap anisotropy, eg: $s$-wave, $p_x+ip_y$ or $d_{x^2+y^2}+id_{xy}$.
Notice that in these systems, when the superconducting gap on the Fermi surface is smaller than the Fermi energy, the band would demonstrate effectively a Mexican hat dispersion as well, just with a reduced spectral weight.
Therefore, it is trivial to understand that other than an overall 1/2 factor related to the weight reduction, the resulting DOS has the same 1D-like divergent behavior at the gap edge.
Thus it would not be very unusual that in these systems, superconductivity can coexist with other symmetry breaking phases.

Finally, for completeness, it is necessary to consider the effect of anisotropy of the dispersion around the Mexican hat.
Such an anisotropy can in principle lift the strict degeneracy at the band edge, effectively recovering the higher-dimensional behavior with less singular DOS.
However, this is obviously a very small energy scale, especially when the radius of the Mexican hat is small.
Above this small energy range, the tendency toward a divergent behavior would still be present, so our above discussion remains valid.
Of course, if one drives the system into the much deeper band inversion phase where the radius of the Mexican hat becomes larger, the anisotropy can be more effective in lifting the degeneracy.
In that case, the system might go through another Lifshitz transition at low doping, from two sheets of Fermi surface to possible multiple pockets.
Fig.~\ref{fig:fig3} (c) demonstrates such a possibility, corresponding to adding anisotropic 3rd order terms in the dispersion relation.
Furthermore, the anisotropy might deform the Fermi surface in ways that would improve the nesting condition for the charge density wave or spin density wave states.
All these interesting possibilities allow further tunability of the bulk physical properties of the generic band-inverted systems.

In summary, we point out that deep into the band inverted state, the topological insulators are generically vulnerable against symmetry breaking instability, due to the novel van Hove singularity near the chemical potential.
This new class of VHS is caused by the characteristic Mexican-hat dispersion at the band edge, which effectively reduces the codimension of the degenerate extrema to one, and guarantees the divergent DOS with a 1D-like exponent for both 2D and 3D cases.
This singular DOS can boost up the instability of the system against a superconducting state and encourages the search for building topological superconductors along this line.
In addition, associated with the formation of the Mexican-hat-like dispersion, a doped system would experience a Lifshitz transition that may multiply the number of Fermi surfaces with modified shapes.
Our study not only explains the existing experimental observations, but also suggests a route to activate additional symmetry breaking phases in the topological insulators, particularly for the long-sought topological superconductivities.

We thank Weiguo Yin, Genda Gu, Liang Fu, Xi Dai, Yaomin Dai, and  Yanliang Shi for very helpful discussions and especially, Christopher C. Homes who brought to us the important experimental results which might support our theoretical predictions of the novel VHS. Work funded by the U. S. Department of Energy,  Office of Basic Energy Sciences DE-AC02-98CH10886.


\begin{thebibliography}{99}
\bibitem{comment} The bare pairing susceptibility ${\chi ^0}\left( {|q| \to{0^ + }, \omega=0} \right)$: ${\chi ^0_{2D}}=\frac{1}{2}g\left( {{\mu}} \right)\ln \left( {\frac{{q{v_F}}}{{4{\omega _D}}}} \right)$ and ${\chi ^0_{3D}}=g\left( {{\mu}} \right)\ln \left( {\frac{{q{v_F}}}{{2{\omega _D}}}} \right)$, is proportional to the density of state at the chemical potential.  Here $v_F$ is the Fermi velocity and $\omega_D$ is the Debye frequency.
This indicates a singularly large DOS near the chemical potential $\mu$ would directly enhance the pairing instability.
\end{thebibliography}
\end{document}